\documentclass[prb,twocolumn,superscriptaddress,showpacs,floatfix]{revtex4}
\usepackage{graphics}
\usepackage{graphicx}
\begin{document}
\newcommand{\RR}{\mathrm{\mathbf{R}}}
\newcommand{\rr}{\mathrm{\mathbf{r}}}
\newcommand{\defin}{\stackrel{def}{=}}
\title{Exchange coupling in semiconductor nanostructures: \\ Validity and
limitations of the Heitler-London approach}
\author{M.J. Calder\'on}
\affiliation{Condensed Matter Theory Center, Department of Physics,
University of Maryland, College Park, MD 20742-4111}
\author{Belita Koiller}
\affiliation{Condensed Matter Theory Center, Department of Physics,
University of Maryland, College Park, MD 20742-4111}
\affiliation{Instituto de F\'{\i}sica, Universidade Federal do Rio de
Janeiro, Caixa Postal 68528, 21941-972 Rio de Janeiro, Brazil}
\author{S. Das Sarma}
\affiliation{Condensed Matter Theory Center, Department of Physics,
University of Maryland, College Park, MD 20742-4111}
\date{\today}
\begin{abstract}
The exchange coupling of the spins of two electrons in double well
potentials in a semiconductor background is calculated  within the Heitler-London (HL)
approximation. Atomic and quantum dot types of confining potentials
are considered, and a systematic analysis for the source of
inaccuracies in the HL approach is presented. For the strongly
confining coulombic atomic potentials in the H$_2$ molecule, the most dramatic
failure occurs at very large interatomic distances, where HL predicts
a triplet ground state, both in 3D and in 2D, coming from the absence
of electron-electron correlation effects in this approach. For a 2D
double well potential, failures are identified at relatively smaller
interdot distances, and may be attributed to the less confining nature
of the potential, leading to larger overlap. We
find that in the double dot case, the range of validity of HL is
improved (restricted) in a related 3D (1D) model, and that results
always tend to become more reliable as the interdot distance
increases. Our analysis of the exchange coupling is of relevance to the
exchange gate quantum computer architectures in semiconductors. 
\end{abstract}
\maketitle
\pagebreak
\section{Introduction}
\label{sec:introduction}
Semiconductor spin-based quantum computation emerged from theoretical
proposals showing that  the required universal gates~\cite{barenco95} could be
implemented through physical operations involving single electrons bound
to an array of quantum dots\cite{Exch,burkard99,HD,sousa01,tarucha96,kouwenhoven97,petta05,johnson05,gorman05} or to donors.\cite{Kane}
The localized spin of each electron serves as the single qubit by
virtue of the 2-level spin-dynamics, and the (electrostatic) quantum
mechanical exchange coupling between the two electrons can be used to
entangle the qubits through the ``exchange gate''.

Accurate calculations of exchange coupling in molecules (or artificial
molecules) are extremely difficult requiring numerically intensive
self-consistent solutions of Hartree-Fock equations. In semiconductor
nanostructures, particularly in the context of quantum computer
architectures, such intensive quantum chemistry-type numerical
calculations are unwarranted for several reasons. First, the parameters
entering such calculations are essentially only very approximately known
in semiconductor nanostructures and, therefore, extremely accurate
quantitative calculations for the exchange energy in these nanostructures
is not particularly meaningful. Second, the quantum chemical type exchange
calculations are not particularly valid in the semiconductor environment.

One of the simplest and most successful methods for the calculation of
exchange coupling in small molecules was proposed in 1927 by Heitler and
London.\cite{heitler27,herring62}  
The basic assumption of the Heitler-London (HL) method is that the
many-electron lowest energy wavefunctions in molecules may be written in
terms of the one-electron ground-state orbitals of the isolated
constituent atoms. This immediately requires that the overlap (S) among
neighboring orbitals be small, a condition not always fulfilled
in real molecules. In fact, for the equilibrium interatomic distance in
the H$_2$ molecule ($1.5\,a_B$, with $a_B$ representing the Bohr radius),
the overlap is larger than 0.7, and the HL ground-state energy is
overestimated by $\sim 1.5$ eV, while for interatomic distances $R \gtrsim
5\,a_B$ the overlap becomes smaller than 0.1 and an excellent agreement is
obtained between HL and the experimentally
observed ground-state energy.\cite{slater} The HL approximation is thus
expected to be valid for $S \ll 1$, or equivalently, for well separated
atoms.

A somewhat surprising anomaly of the HL method applied to H$_2$ is that it
predicts a triplet ground-state for $R \gtrsim 50\, a_B$, in contradiction with the well established result that, for a spin-independent Hamiltonian and in the absence of magnetic fields, the ground-state of an even-electron system in a symmetric potential must be a
singlet.\cite{lieb62} This anomaly in HL comes from the singular nature of the Coulomb potential,\cite{herring62} leading to a logarithmic dependence of the electron-electron repulsion energy with $R$ which becomes the dominant term at large $R$, resulting in a negative value of the exchange coupling $J(R) = E_{triplet} - E_{singlet}$ in this limit.  An asymptotically correct expression for  $J(R)$ for the hydrogen molecule was obtained by Herring and Flicker,\cite{herring64} who have also shown that the HL results for     $J(R)$ are in excellent quantitative agreement with the asymptotic expression up to $R$ values very near the crossover to the unphysical negative range of $J(R)$.

Many calculations of exchange coupling in semiconductor quantum dots are
based in the HL approximation applied to 2D model
potentials. It has been pointed out that, for some parameter values of a
quartic model potential proposed for the study of 2D double quantum dots,
unphysical (negative) values of $J$ are obtained within the HL
approximation.\cite{burkard99} This failure of HL may be due to the
reduced dimensionality of the potential, since the electron-electron
Coulomb interaction in 2D is expected to be larger than in 3D, or to some
intrinsic limitation of this approach when applied to weakly confining
gated potentials as compared to atomic-like molecular potentials.
We address these questions here by considering the 2D hydrogen
molecule
and the two-electron double
quantum dot quartic potential within HL, including its modified
versions in 3D and 1D.

Our study is complementary to detailed and numerically intensive
calculations of exchange. We find that the reliability of HL
depends primarily on the form of the confining potential: (i) For weakly
confining (low  barrier) potentials, the range of validity is
generally improved in larger space dimensions; (ii) For strongly
confining potentials, illustrated here by the coulombic potential, the
space dimensionality is not a decisive parameter; (iii) Unreliable and
even unphysical results may be attributed to non-additive model
potentials intended to simulate a double-well environment. 

This paper is organized as follows: In Sec.~\ref{sec:hydrogen} we
quickly review the HL method for the H$_2$ molecule, and 
consider the two-dimensional analogue of this problem; in
Sec.~\ref{sec:quartic} we present the HL solution for the two-electron
problem in the 2D double quantum dot quartic potential introduced in
Ref.~\onlinecite{burkard99} and compare it to its modified versions in
3D and in 1D. Further discussions and conclusions are presented in
Sec.~\ref{sec:conclusions}. 

\section{The HL method for the H$_2$ molecule in 2D and 3D}
\label{sec:hydrogen}
The Hamiltonian of the hydrogen molecule with nuclear coordinates ${\bf R}_A$ and ${\bf R}_B$ is written as\cite{slater}
\begin{equation}
H =
T_1 + T_2 + V({\bf r}_1) + V({\bf r}_2) + {e^2\over r_{12}}+{e^2\over R}~,
\label{eq:h}
\end{equation}
where $V({\bf r}_i)= V_A({\bf r}_i)+V_B({\bf r}_i)= -{e^2\over
r_{ia}}-{e^2\over r_{ib}}$, $r_{ia}=|{\bf r}_i-{\bf R}_A|$, $r_{ib}=|{\bf
r}_i-{\bf R}_B|$, ${\bf r}_i~(i=1,~2)$ are the electronic coordinates,
$T_i$ is the kinetic energy operator for electron $i$, $r_{12} = |{\bf
r}_1-{\bf r}_2|$ and $R=|{\bf R}_A-{\bf R}_B|$ is the inter-nuclear
separation. 
Starting from normalized hydrogen atomic orbitals centered at
${\bf R}_A$ and ${\bf R}_B$, which we denote for each electron $i$ as
$a(i)$ and $b(i)$, the HL lowest singlet $(+)$ and triplet $(-)$
2-electron states (we omit the spin-dependent part here) are written as
\begin{equation}
|\pm\rangle = {1\over \sqrt{2(1\pm S^2)}}[a(1)b(2)\pm b(1)a(2)]~,
\label{eq:hl}
\end{equation}
with $S=\langle a(i)|b(i) \rangle$ giving the overlap integral.
It is convenient to cast the exchange coupling, $J= E_{triplet} -
E_{singlet} \cong \langle - |H| - \rangle - \langle + |H| + \rangle$ in
the following form
\begin{equation}
J = {2 S^2 \over 1 - S^4} (W - C)
\label{eq:j}
\end{equation}
where
\begin{equation}
W=\langle a(1)b(2) |v| a(1)b(2) - {1\over S^2} a(2)b(1)\rangle~
\label{eq:w}
\end{equation}
with  $v=V({\bf r}_1) + V({\bf r}_2) - V_B ({\bf r}_2) - V_A({\bf
r}_1)$, and
\begin{equation}
C=\langle a(1)b(2) |{e^2\over r_{12}}| -a(1)b(2) + {1\over S^2}
a(2)b(1)\rangle~.
\label{eq:c}
\end{equation}
The two terms  $W$ and $C$ are always positive, and correspond to
energy terms which contribute differently to the singlet and triplet
states: $W$ is related to the ``covalent'' energy, which favors the
singlet state providing molecular bonding, while $C$ is related to the
electron-electron Coulomb repulsion, favoring the triplet state. Note that both terms are independent of the kinetic energy. As discussed in the introduction, physically $J>0$, therefore we expect $W>C$.

The HL solution was presented long ago for the 3D case,~\cite{heitler27}
where an isolated atomic orbital has the form
$
a_{3D}(i)= e^{-r_{ia}}/\sqrt{\pi}~,
$
and equivalently for $b_{3D}(i)$.
Distances are given in units of $a_B=\hbar^2/(me^2)$.
The 3D overlap is
$
S_{3D}= \exp(-R) \left(1+R+R^2/3\right)~.
$

The 2D version of this problem has not been previously discussed within HL. We present it here since it is useful in addressing the effects of dimensionality, particularly regarding the electron-electron Coulomb potential. The ground-state orbital for the 2D hydrogenic atom is\cite{ponomarev991,ponomarev992}
$a_{2D}(i)=4 \exp(-2 r_{ia})/\sqrt{2 \pi}~,$
leading to the overlap
$S_{2D}=2R^2 \left[K_0(2R)+K_1(2R)/R \right]~,$
where $K_0$ and $K_1$ are modified Bessel functions of the second kind.
The problem in 2D is solved following the same procedure as in 3D, in
particular using a 2D version of the spheroidal coordinates
$\lambda=(r_a+r_b)/R$ and $\mu=(r_a-r_b)/R$ described
in detail in Ref.~\onlinecite{slater}. Contrary to the 3D case, for which analytic expressions are obtained for all terms, in 2D the electron-electron Coulomb interaction terms contributing to $C$ in Eq.~(\ref{eq:c}) are obtained numerically.

\begin{figure}
\begin{center}
\resizebox{70mm}{!}{\includegraphics{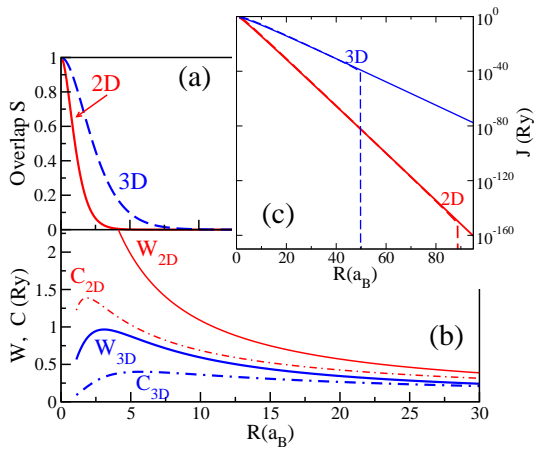}}
\caption{\label{fig:hydrogen}(Color online) (a) Overlap between the
hydrogen ground-state orbitals centered at nuclear sites a distance
$R$ apart in 2D and 3D. (b) Covalent (W) and Coulomb (C) energy terms
contributing to the electronic exchange within the HL approach in the
hydrogen molecule in 2D and 3D, in units of $Ry$. The downturn region
at short distances ($R<4$) is less reliable due to the large
overlap. (c)  The solid lines are the
asymptotic values for the electronic exchange for the hydrogen
molecule in 2D and 3D,~\cite{herring64,ponomarev991} $J_{2D}^{\rm asymptotic}= 15.2 R^{7/4} \exp(-4R)$ and
$J_{3D}^{\rm asymptotic}= 0.8 R^{5/2} \exp(-2R)$, that we
compare here with the Heitler-London results (dashed lines). The
comparison is very good up to the large distances where HL fails. The
HL results are dominated by the prefactor $S^2$, see Eq.(\ref{eq:j}),
except for the
negative-$J$ anomaly, indicated here by the steep drop in the HL
curves.
}
\end{center}
\end{figure}

In Fig.~\ref{fig:hydrogen} we present the results for the overlap
($S$), the covalent energy ($W$), the Coulomb energy ($C$), and the
exchange ($J$) in two and three dimensions. Energies are given in units of Ry$=m e^4/(2 \hbar ^2)$. The first thing that shows up is that, although the overlap $S_{2D}(R)$ is smaller than $S_{3D}(R)$ for all $R$, both $W$ and $C$ contributions, as well as their difference $W-C$, are larger in 2D than in 3D, which reflects the enhanced Coulomb interaction due to the electronic charge additional confinement in 2D as compared to 3D.
Fig.~\ref{fig:hydrogen}(c) illustrates the well known
artifact of HL for the 3D hydrogen problem\cite{herring64} giving
negative 
$J$ at
very large distances (above $R\sim 50 a_B$). At shorter distances, the comparison of $J$ given by Eq.~(\ref{eq:j}) with the most accurate asymptotic results available in the literature for 2D and 3D~\cite {herring64,ponomarev991} 
is very good for a wide range of distances.
Within our numerical accuracy, in 2D the critical $R$
at which $J$ becomes negative is between $40$ and $100 a_B$, comparable to the 3D value.
At very short distances, HL becomes less reliable due to the large overlap
[see Fig.~\ref{fig:hydrogen}(a)]. Arbitrarily defining a cutoff at
$S=0.1$, we estimate the minimum distance at which we can trust the HL
results to be $R_{min}^{3D} \sim 5 a_B$, and $R_{min}^{2D} \sim 2.5 a_B$, since the overlap for the 2D system decays much faster with $R$ than for 3D.

\section{The HL method for the quartic double-dot potential}
\label{sec:quartic}
Quantum dots are defined by  gate-generated potentials modifying a two-dimensional electron gas (2DEG) environment in semiconductor heterostructures.
Each quantum dot is usually modeled by a  harmonic well with $V_j({\bf r})=
{{m\omega_o^2}\over{2}}({\bf r}-{\bf R}_j)^2$ ~ where  ${\bf R}_j=+R/2\, {\hat x}$, $-R/2\, {\hat x}$ ~ for $j=A,B$ respectively.
Following Burkard {\it et al.},~\cite{burkard99} the 
coupling of the dots in a 2D system is modeled by the quartic potential
\begin{equation}
V(x,y)={{m\omega_o^2}\over{2}}
\left\{{\left[{x^2-\left({{R}\over{2}}\right)^2}\right]^2\over{R^2}}+y^2\right\}
\label{eq:v2d}
\end{equation}
which satisfies $V(x \approx \pm R/2,y)= V_j({\bf r})$. Note that the
barrier height increases with $R$.
The Hamiltonian for two electrons in a double quantum dot can be written
as
\begin{equation}
H=T_1+T_2+V({\bf r}_1)+V({\bf r}_2)+{{e^2}\over{\epsilon r_{12}}}
\end{equation}
where $\epsilon$ is the semiconductor dielectric constant. 
For the results presented below, distances are given in units of $a^*=\hbar^2
\epsilon/(m^*e^2)$ and energies in Ry$^*=m^* e^4/(2 \hbar^2
\epsilon^2)$, where $m^*$ is the effective mass in the host
semiconductor.
Note that, contrary to the H$_2$ case, $V \ne V_A + V_B$.
The single electron Hamiltonian is
\begin{equation}
h_i=T_i+V_A({\bf r}_i)~,
\end{equation}
the corresponding ground state orbital 
is $a_{2D}(i)={\beta\over \sqrt{\pi}}\, \exp \left\{-{{\beta^2}\over{2}}
\left[\left(x_i \pm {{R}\over{2}}\right)^2+y_i^2\right]\right\}$, with
$\beta=\sqrt{{m^* \omega_0}/{\hbar}}$, and the overlap is given by $S_{2D}=\exp(-\beta^2 (R/2)^2)$.

We point out that the physical significance of the covalent term $W$ defined in Eq.~(\ref{eq:w}) is somewhat different here as compared to the H$_2$ system. From Eq.~(\ref{eq:w}), the operator
involved in $W$ is an effective potential $v=V({\bf r}_1) + V({\bf r}_2) - V_B ({\bf r}_2) - V_A({\bf r}_1)$ which, for H$_2$, results in $v=V_B ({\bf r}_1) + V_A({\bf r}_2)$, related to the energy of an ``$A$-atom electron'' due to the potential $V_B$ of the $B$-atom and vice versa. In the double quantum dot, $v \ne V_B ({\bf r}_1) + V_A({\bf r}_2)$, instead 
the effective potential $v$ involves differences between the quartic potential and  
quadratic terms. In this way, the ``electron in the $A$-dot'' does not ``feel'' the potential $V_B$, but an approximated version of it.

Using the expressions for $W$ and $C$ given in Eqs.~(\ref{eq:w}) and (\ref{eq:c}) we get,~\cite{burkard99}
\begin{equation}
W_{2D}={3\over 4}\hbar \omega_0 \left(1+{{\beta^2
R^2}\over{4}}\right)~,
\label{eq:w2D}
\end{equation}
and
\begin{equation}
C_{2D}=\hbar \omega_0\, c \left[1-e^{-{{\beta^2 R^2}\over{4}}}
I_0\left({{\beta^2 R^2}\over{4}}\right)\right]
\label{eq:c2D}
\end{equation}
where $I_0$ is the modified Bessel function of the first kind and
\begin{equation}
c=\sqrt{\pi \over 2}\, {{e^2
\beta/\epsilon}\over{\hbar \omega_0}}~
\label{eq:cbld}
\end{equation}
is a parameter introduced in Ref.~\onlinecite{burkard99} as the ratio between the Coulomb and the confining energy.
Note that  $c$ is also completely
defined by $\omega_0$, since $\beta \sim \sqrt{\omega_0}$. 

In order to investigate the effect of dimensionality, it is instructive
 to compare the 2D system with a similar
3D problem. From the 2D potential in Eq.~(\ref{eq:v2d}), we define a 3D counterpart as $V_{3D}(x,y,z)=V(x,y)+{{m\omega_o^2}\over{2}}z^2 $,
with the ground state orbital  $a_{3D}(i)={\beta^{3/2}\over {\pi^{3/4}}} \exp{\left\{-\beta^2
\left[\left(x_i \pm {{R}\over{2}}\right)^2+y_i^2+z_i^2\right]/2\right\}}$,
 leading to $S_{3D}=S_{2D}$, $W_{3D}=W_{2D}$ and
\begin{equation}
C_{3D}={{2}\over{\pi}}\hbar \omega_0\, c \left[1-\sqrt{\pi \over 2}\,
{{{\rm Erf}(\beta R/\sqrt{2})}\over{\beta R}} \right]~,
\label{eq:c3D}
\end{equation}
where ${\rm Erf}(x)={2\over\sqrt{\pi}} \int_0^x e^{-u^2}\, du$ is the
Error Function.

\begin{figure}
\begin{center}
\resizebox{70mm}{!}{\includegraphics{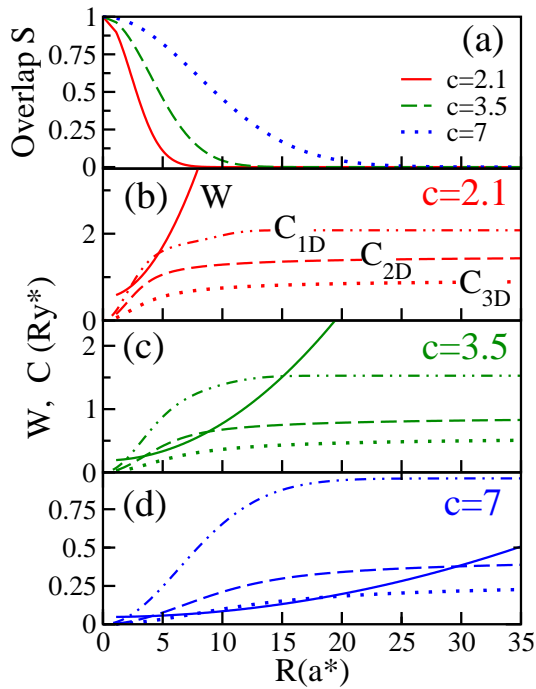}}
\caption{\label{fig:quartic}(Color online) (a) Overlap ($S=S_{1D}=S_{2D}=S_{3D}$) for three
different values of the parameter $c$ [defined in Eq.~(\ref{eq:cbld})] for the
double quantum dot problem. (b), (c), and (d) Covalent ($W=W_{1D}=W_{2D}=W_{3D}$)
and Coulomb ($C$) energy terms for the double quantum dot problem in 1D, 2D, and
3D. When the overlap is large, HL results are not reliable and, for larger values
of $c$, lead to negative exchange $J$.} 
\end{center}
\end{figure}

For completeness, we also discuss the 1D problem. The model potential is  $V_{1D}(x)={{m\omega_o^2}\over{2}}
\left[x^2-\left({{R}\over{2}}\right)^2\right]^2/R^2$. 
The single-electron ground state is 
$a_{1D}(i)={\sqrt{\beta}\over \pi^{1/4}} \exp \left\{-\beta^2 \left(x_i \pm {{R}\over{2}}\right)^2/2\right\}$,
 leading to $S_{1D}=S_{2D}$, $W_{1D}=W_{2D}$. The Coulomb term corresponds to the $\varepsilon=0$ value of the function
\begin{eqnarray}
& &C_{1D}(\varepsilon)={{2}\over{\pi}}\hbar \omega_0\, c\,\int_{\varepsilon}^\infty dx\,
{{e^{-\beta^2 x^2/2}}\over{x}} \nonumber\\
&-& {{2}\over{\pi}}\hbar \omega_0\, c\, S^2 \,
\int_{\varepsilon}^{\infty}dx \,{{e^{-\beta^2 x^2/2} \cosh(\beta^2 R x)}\over{x}}~,
\label{eq:c1D}
\end{eqnarray}
where $x=|x_1-x_2|$.
In the $\varepsilon \rightarrow 0$ limit, the singular nature of the Coulomb potential in 1D systems leads to a logarithmic divergence of this expression. We overcome this problem phenomenologically by introducing a cutoff distance (the results presented in Fig.~\ref{fig:quartic} correspond to $\varepsilon= 0.2 a^*$) to simulate the correlation effects that would
lead to a lower probability density at $x_1=x_2$. Such correlation effects are not included in HL.\cite{herring62}

Fig.~\ref{fig:quartic} shows the behavior of $S$, $W$ and $C$ vs $R$
for different values of $c$ and dimensionalities.  The HL approach breaks down for small $R$ for a certain
value of $c$ which depends on dimensionality. For 2D, the breakdown
occurs when $c>2.8$, as mentioned by Burkard {\it et
al.}~\cite{burkard99} Meanwhile, for 3D HL fails for $c>5.8$, and for
$c>1.95$ for 1D.~\cite{foot-1D}
On the other hand, at large enough $R$ one always gets $W>C$ and thus positive $J$, regardless of
$c$. Hence, the large-$R$ anomaly encountered in
the H$_2$ HL solution does not occur here.
We may argue that the large-$c$ failure of the HL approximation  is ultimately due to a
large-$S$ regime [see Fig.~\ref{fig:quartic}(a)].

\begin{figure}
\begin{center}
\resizebox{70mm}{!}{\includegraphics{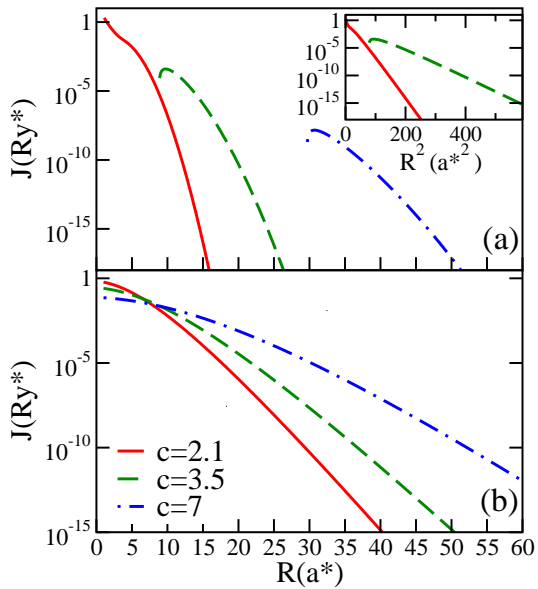}}
\caption{\label{fig:quarticJ}(Color online) (a) Exchange at long distances
$R$, within HL approximation,  in the two-dimensional double quantum dot
system, for three different values of $c=2.1$, $3.5$, and $7$. For the
larger values of $c$, HL breaks down at small $R$. The inset displays the
curves for $c=2.1$ and $3.5$ plotted versus $R^2$, to show that, at
large $R$, the decay of $J$
is dominated by the $S^2$ behavior (see text). (b)
Exchange as calculated by Ponomarev {\it et al.} (Eq. 44 in
Ref. \onlinecite{ponomarev992}) for the same values of $c$.}
\end{center}
\end{figure}

The exchange calculated from Eq.~(\ref{eq:j}) for the 2D system is given in 
Fig.~\ref{fig:quarticJ} (a) for the same values of $c$ as in Fig.~\ref{fig:quartic}.  
Note that for the largest $c$ values the curves are interrupted when
$J$ becomes negative, and for $R$ just above this point HL predicts
the equally unphysical behavior of $J$ {\it increasing} with
$R$.~\cite{foot} For $c=2.1$, $J$ decreases monotonically with $R$,
but the ``shoulder'' at $R\sim 4a^*$ is a precursor signaling this
anomaly, meaning that $J>0$ is not a sufficient condition for the
reliability of the HL results.  For distances far beyond these
anomalous points, we find that the general behavior of $J(R)$ is
dominated by the $S^2$ prefactor in Eq.~(\ref{eq:j}), as shown in the
inset to Fig.~\ref{fig:quarticJ} (a).

From a formal point of view, it would be interesting to present a
direct comparison of the $R\to\infty$ behavior of $J(R)$ calculated 
within HL with asymptotic expressions such as given in
Fig.~\ref{fig:hydrogen}(c) for H$_2$, obtained through the 
so-called median-plane method. This method was originally introduced
for two hydrogen atoms,\cite{herring64} and has been extended 
to a variety of problems in atomic and semiconductor
physics.\cite{gorkov03} It is however not directly applicable to the
potential in Eq.~(\ref{eq:v2d}) because, in 
the asymptotic limit, this model potential displays an essentially
infinite barrier between the dots.\cite{foot1}     
We present instead a qualitative comparison of our results in
Fig.~\ref{fig:quarticJ}(a) with those obtained from an interpolated
formula for the exchange coupling of two electrons under a potential
related to ours,\cite{ponomarev992} namely the potential created by
donor impurities close to a 2DEG.  
The connection between our results and those in
Ref.~\onlinecite{ponomarev992} is based on the fact that  
when a single impurity is far from the 2DEG, the
electronic potential it creates may be approximated by a parabolic
confinement, whose curvature defines $\omega_0$ and thus $c$. The
results plotted in Fig.~\ref{fig:quarticJ}(b) are obtained from
Eq.(44) in Ref.~\onlinecite{ponomarev992} for distances between the
donor and the 2DEG $h=2$, $4$, and $10 a^*$, resulting in $c=2.1$, 3.5
and 7 respectively. We note that the general trends in
Figs.~\ref{fig:quarticJ}(a) and (b) are the same, though a detailed
quantitative agreement is not found. This lack of agreement is due to
the different nature of the barrier between the parabolic potentials. 
As noted above, the exchange energy
within HL scales asymptotically as $S^2$, thus for the quartic model
the exchange has an exponential decay with $R^2$, while for the impurity
potential in Ref.~\onlinecite{ponomarev992} the two-electron
wave-functions, and consequently $S$ and $J$, have a slower 
exponential decay with $R$. 

We conclude that HL is essentially reliable, at least qualitatively, for the double quantum dot
problem at large enough interdot separations.  
In the large overlap region, HL is not applicable
regardless of the sign of $J$.  
We have attempted a standard improvement to HL, which for 
the H$_2$ problem is to consider the Bohr radius as a variational
parameter.~\cite{slater} In the present case this consists in taking
$\beta$ as a variational parameter: Although we obtain some lowering
of the singlet and triplet energies by this procedure, no qualitative 
improvement is obtained.
Another straightforward improvement is the Hund-Mulliken approach~\cite{burkard99} in which the double occupancy states $a(1)a(2)$ and
$b(1)b(2)$ are included in the basis set for the 2-electron states. This method lowers the singlet energy with respect to the triplet, but the failure at large $c$ remains.

\section{Summary and Conclusions}
\label{sec:conclusions}
In this work, we have analyzed the reliability of HL for two
types of ``double-well'' potentials where the individual-well
confinement potentials are of different natures: Atomic (Hydrogen
atom) in Sec.~\ref{sec:hydrogen} or harmonic (quantum dot) in
Sec.~\ref{sec:quartic}. There are two main differences between these
problems: One refers to the strength of the confinement, much stronger
in the first case, and the other to their dimensionality, 3D for atoms and 2D for quantum dots defined over a 2DEG.  
 
For the 3D H$_2$ case, it has been long known~\cite{herring64} that HL fails, namely, the triplet state is predicted to become the ground-state, when the atoms are very far apart ($R\sim 50 a_B$). As argued above, at short distances HL is expected to fail as well, and it does as it overestimates the ground-state energy. We have also considered the 2D hydrogenic molecule within HL, and found qualitatively the same behavior and limitations as for the 3D H$_2$ case, meaning that the strongly confining atomic potential is the dominant aspect of this problem both in 3D and in 2D.

Our study of the particular quartic potential proposed by Burkard
{\it  et al.}\cite{burkard99} for modeling coupled harmonic dots shows
that the failure of HL at short distances is more dramatic here than
in H$_2$ due to the less  confining potential, leading to larger
overlap: A triplet ground-state is predicted for a wide range of
parameters. We have also pointed out the fact that the quartic
potential is not additive (i.e. it does not correspond to the
superposition of the isolated harmonic potentials) as another possible
source for this failure. Overcoming these limitations for the exchange calculation, in
particular when the overlap is large, requires going beyond the HL
approximation, for example increasing the basis set for the two- or
many-electron wave-functions to incorporate a set of excited
one-electron states.  Different methods have been adopted in the
literature: Molecular orbital, \cite{HD,sousa01}
configuration-interaction,\cite{HD1} exact diagonalization,
\cite{scarola05,helle05} and unrestricted Hartree-Fock.\cite{yannouleas02}
Molecular orbital calculation of double-dot model potentials have been
explicitly compared to HL,\cite{burkard99,sousa01} and in all cases an
excellent qualitative agreement is obtained. According to the model
potential, practically complete agreement may be obtained,\cite{burkard99}
while deviations of up to 50\% have also been reported.\cite{sousa01}    
Also, spin polarized DFT calculations can be done to obtain the
exchange energy,\cite{Anisimov,stopa05} but it is difficult to
estimate the reliability of such theories in the present 
scenario. Unfortunately none of the cited methods has the conceptual and
computational simplicity of HL, which, coupled to a careful choice for
the model potential, should be always attempted as a first
approximation for estimating exchange coupling in new systems.  

\begin{acknowledgments} 
We thank Pavel Krotkov for useful discussions.
This work is supported by LPS and NSA. BK
also acknowledges support by CNPq, FUJB, Millenium Institute-CNPq, and
FAPERJ.
\end{acknowledgments}
\bibliography{exchange}

\begin{thebibliography}{29}
\bibitem[{foo({\natexlab{a}})}]{foot-1D}
\bibinfo{note}{This is the case for the particular cutoff considered
  ($\varepsilon= 0.2 a^*$). The smaller the cutoff parameter $\varepsilon$, the
  larger the value of $C_{1D}$, and the lower the value of $c$ that will cause
  HL to give a negative exchange. Strictly speaking, HL always predicts a
  triplet ground state for this 1D model, a characteristic that should be
  expected in 1D systems if electron-electron correlations are not taken into
  account.}
\bibitem[{foo({\natexlab{b}})}]{foot}
\bibinfo{note}{Band interference effects, which are not relevant here, may lead
  locally to this behavior as obtained in Ref.~\onlinecite{KHD1}.}
\bibitem[{foo({\natexlab{c}})}]{foot1}
\bibinfo{note}{The median-plane method consists of reducing the expression for
  $J$ to a surface integral over a hyperplane in coordinate space, and using a
  perturbative approach to determine the two-electron wavefunction. The
  accuracy of the perturbative approach has been demonstrated (see for example
  Refs.~\onlinecite{herring62} and \onlinecite{herring64}) for additive
  potentials such that the contribution from each confining center tends to
  zero at large distances from the location of its minimum.}
\end{thebibliography}
\end{document}